\documentclass[preprint]{aastex}

\newcommand{\kms}{km s$^{-1}$}

\shorttitle{On the distance of Magellanic Clouds}
\shortauthors{Bono et al.}

\begin{document}

\title{On the distance of Magellanic Clouds: first overtone Cepheids}

\author{G. Bono\altaffilmark{1}, M.A.T. Groenewegen\altaffilmark{2}, 
M. Marconi\altaffilmark{3}, F. Caputo\altaffilmark{1}}  

\affil{1. Osservatorio Astronomico di Roma, Via Frascati 33,
00040 Monte Porzio Catone, Italy; bono/caputo@mporzio.astro.it}
\affil{2. Instituut voor Sterrenkunde, PACS-ICC, Celestijnenlaan 200B,
3001 Heverlee, Belgium; groen@ster.kuleuven.ac.be}  
\affil{3. Osservatorio Astronomico di Capodimonte, Via Moiariello 16,
80131 Napoli, Italy; marcella@na.astro.it}

\begin{abstract}
We present a detailed comparison between predicted and empirical
$PL_{I,K}$ relations and Wesenheit function for Galactic and
Magellanic Clouds (MCs) First Overtone (FO) Cepheids. We find that 
zero-points predicted by Galactic Cepheid models based on a noncanonical 
(mild overshooting) Mass-Luminosity (ML) relation are in very good 
agreement with empirical zero-points based on HIPPARCOS parallaxes, 
while those based on canonical (no overshooting) ML relation
are $\approx 0.2-0.3$ mag brighter. We also find that predicted and
empirical $PL_K$ relation and Wesenheit function give, according to
optical ($V,I$ OGLE) and near-infrared (NIR, $K$, {\sc 2mass}) data,
mean distances to the MCs that agree at the 2\% level. 
Individual distances to the Large and the Small Cloud are: 
$18.53\pm0.08$-$19.04\pm0.11$ (theory) and 
$18.48\pm0.13$-$19.01\pm0.13$ (empirical). 
Moreover, predicted and empirical FO relations 
do not present, within the errors, a metallicity dependence.
Finaly, we find that the upper limit in the FO period distribution 
is a robust observable to constrain the accuracy of pulsation 
models. Current models agree within 0.1 in $\log P$ with the 
observed FO upper limits.

\end{abstract}


\section{Introduction} 
 
The massive photometric databases collected by the micro-lensing
experiments (MACHO, EROS, OGLE) substantially increased the number of
known variable stars in the Galaxy and in the MCs. They also provided 
the unique opportunity to improve the sampling along the light curves 
of fundamental (F) and overtone pulsators. In particular, the new 
multi-band data on classical Cepheids have had a substantial impact 
not only on the pulsation properties of
these variables but also on the estimate of MC distances (Udalski
1998; Groenewegen \& Oudmaijer 2000, GO00; Groenewegen 2000, G00; 
Bono, Caputo, \& Marconi 2001, BCM01).
Even though F Cepheids are the most popular standard
candles to estimate distances, several theoretical and empirical
investigations have been recently focused on Cepheids pulsating in
the first (G00; Feuchtinger, Buchler \& Koll$\acute{a}$th 2000, FBK00; 
Baraffe \& Alibert 2001, BA01) or in the second overtone (BCM01).  
The main advantage in using overtone
pulsators to estimate distances is that the width in temperature 
of their instability regions is significantly smaller than for the
fundamental one. As a matter of fact, current predictions suggest 
that at $\log P=0.3$ the width of FO instability strip is 400 K, 
while for F variables it is 900 K at $\log P=1$. Therefore, distances 
based on FO Period-Luminosity (PL) relations are marginally affected 
by intrinsic spread when compared with F ones. However, we still 
lack a comprehensive analysis of the uncertainties affecting distance
estimates based on optical and NIR PL relations and on the Wesenheit 
function. A similar approach was already adopted by
G00 but in the comparison between theory and observations he was
forced to fundamentalize the period, since theoretical predictions 
for FO in MCs were not available. This gap was filled by BA01 who 
estimated PL, PLC relations, and Wesenheit function on the basis 
of linear, convective models. 
Although FO variables present several undisputable advantages, 
complete and accurate samples are only available for MC Cepheids.
Moreover, the detection of FOs in external galaxies is more difficult 
than for F Cepheids, since they are fainter and the luminosity amplitudes 
are smaller.

The main aim of this Letter is to give a detailed analysis of pros
and cons of the PL relations currently adopted to estimate the 
distances, according to full amplitude, nonlinear, 
convective models of Galactic and MC Cepheids. We are also interested 
in testing whether current models account for the observed upper limit 
in the FO period distribution.

\section{Comparison between theory and observations}

To investigate the topology of the instability strip of Galactic and
MC Cepheids, pulsation models were constructed by adopting three different 
chemical compositions, namely Y=0.25, Z=0.004 (Small 
Magellanic Cloud, SMC), Y=0.25, Z=0.008 (Large Magellanic Cloud, LMC), 
and Y=0.28, Z=0.02 (Galaxy). To assess whether the FO instability strip 
presents a change in the slope when moving from
long- to short-periods we adopted several stellar masses.
For each mass value and chemical composition, the luminosities were
fixed according to the same Mass-Luminosity (ML) relation adopted in
previous investigations (Bono, Marconi \& Stellingwerf 1999, BMS99)
which is consistent, within the errors, with the ML relation
derived by Bono et al. (2000). To mimic the luminosities predicted by
evolutionary models that account for mild convective core overshooting
during H-burning phases (noncanonical) the luminosities predicted by
canonical models were increased by 0.25 dex (Chiosi, Wood \& Capitanio
1993).  Recent evolutionary predictions  
by Girardi et al. (2000) support such a scaling. This choice was
driven by the fact that Cepheid models constructed using the same 
theoretical framework and a noncanonical ML relation account for
the luminosity variation over the pulsation cycle of two LMC bump
Cepheids (Bono, Castellani, \& Marconi 2002). The same outcome does 
not apply to the models based on a canonical ML relation.

The models at solar chemical composition are the ones by Bono et
al. (2001). As for the other chemical compositions, we constructed 
5 new FO sequences with stellar masses ranging from $M/M_{\odot}=3.0$, 
to 5.6 and stellar luminosities ranging from $\log L/L_{\odot}=2.58$, 
to 3.60 to constrain the instability strip in the short-period range. 
The input physics and the physical assumptions 
adopted to construct the models are discussed in BMS99. 
For each model the stability analysis was only performed for the first 
two modes. The physical structure of the static envelope was forced
out of equilibrium by perturbing the radial eigenfunction with a
velocity of 5 \kms. Each case was integrated in time till the
similarity over consecutive cycles was of the order of $10^{-4}$ or
smaller. Close to the instability edges the temperature step is of 100 K.
Bolometric light curves were transformed into the observational plane 
using the atmosphere models by Castelli et al. (1997). The linear 
regressions to estimate PL relations and the reddening-free 
Wesenheit function were performed at fixed chemical 
composition. We adopted this approach because previous relations 
are affected by different systematic uncertainties. The split of the 
FO sample into three sub-samples (Galactic, MCs) reduces the intrinsic 
spread of empirical relations. Therefore the comparison between theory and 
observations should allow us to single out the dependence on metallicity, 
and/or reddening.  

In Table~1 theoretical slopes and zero-points of various
$PL$-relations are listed for canonical and noncanonical FO pulsation
models at solar metallicity, and compared to empirical zero-points
based on HIPPARCOS data. Note that observed zero-points have been
evaluated using the theoretical slopes.  It is recalled that the
accuracy of the HIPPARCOS data for Cepheids was not sufficient to
determine both zero-point and slope of the Galactic $PL$-relation. 
However, for an adopted slope, it is possible to derive in a bias-free
manner the zero-point using the method of ``reduced parallax'' (Feast
\& Catchpole 1997; GO00).  Here, for the first time, HIPPARCOS data
will be used to derive the zero-point of FO $PL$-relations 
adopting the same procedure.  Note however that the periods need 
still to be ``fundamentalized'' to apply the appropriate reddening 
corrections (see GO00 for details). 
The results listed in Table~1 show that predicted noncanonical FO
zero-points are in good agreement with empirical ones both in optical
and NIR bands. On the other hand, FO zero-points predicted by
canonical models are 0.2-0.3 mag systematically brighter than
empirical ones. 
A similar test was also performed using F pulsators and we found
that the difference between predicted noncanonical and empirical
zero-points ranges from 0.08 (W), 0.2 ($PL_K$), to 0.6 ($PL_V$).
It is not clear whether this discrepancy is due to problems in the
reddening correction or to a difference in the period cut-off.
However, we did not find
a systematic difference between canonical and noncanonical
zero-points. This suggests that FO variables are more robust to
constrain theoretical models, however more extended calculations 
are required to constrain the difference. This finding strengthens 
the result obtained by Bono et al. (2002) concerning the accuracy 
of nonlinear, convective, bump Cepheid models. 
They found that theoretical models constructed by adopting a canonical 
ML relation do not account for the luminosity change over the pulsation 
cycle of two LMC bump Cepheids.

To provide independent MC distance evaluations we performed a
linear regression over theoretical models for $Z$=0.008, 0.004.
Table~2 gives zero-points, slopes, and distances obtained using 
theoretical relations.  
The MC FO sample is based on OGLE ($V$ and $I$) and on {\sc 2mass}
($K$) data, reddening corrections (see G00 for details) come from OGLE
estimates (Udalski et al. 1999; Bersier 2000).
The empirical zero-point is derived by fixing the slope of the
$PL$-relation to the theoretical one. Figs.~1 and 2 show that the slopes 
found by fitting the data alone differ slightly from the theoretical 
slopes. However, this is allowed for by
the larger error-bars in the empirical zero-point. Note that lower-limits
in period are enforced to avoid problems with a possible nonlinear 
$PL$-relation at short periods in the SMC and the effect of Malmquist 
bias (see G00). We applied 3$\sigma$ clipping.

The results listed in Table~2 clearly show that the true LMC
distances based on different distance indicators agree quite well and
range from $18.42\pm0.17$ ($PL_I$) to $18.54\pm0.09$ ($W$). The small
discrepancy between the distance based on $PL_I$ relation and the ones
based on $PL_K$ and $W$ function could be due to a mild overestimate
of interstellar reddening. In fact, both $PL_K$ and $W$ function are
only marginally affected by reddening corrections. The same outcome
applies to the SMC and indeed the evaluated distance ranges from
$18.98\pm0.17$ ($PL_I$) to $19.06\pm0.13$ ($W$). However, it is
noteworthy that the relative distance between the clouds attains the
same value, namely 0.51 ($PL_K$), 0.54 ($W$), and 0.56 ($PL_I$) and
agree with relative distances provided by Udalski et al. (1999) and by
G00. Moreover, and even more importantly, these distances are in good
agreement with empirical distances based on observed MC slopes and
zero-points. Note that the "equivalent Galactic zero-points" have been
estimated using HIPPARCOS FO data and the empirical MC slopes.  The
data listed in Table~3 show that empirical solutions provide LMC
distances that range from $18.28\pm0.16$ ($PL_I$) to $18.45\pm0.35$
($PL_K$) and $18.48\pm0.14$ ($W$). At the same time, SMC distances
range from $18.91\pm0.17$ ($PL_I$) to $19.02\pm0.14$ ($W$). 
Note that the difference in the slope between LMC and SMC FOs is due
to the adopted period cutoff. 
The distance moduli obtained in G00 for F pulsators using the same 
method are typically 0.1 larger.
Once again, absolute distances based on the $PL_I$ relation attain
smaller values when compared with distances based on the $PL_K$
relation and on the $W$ function.  As a whole, we find that the
weighted mean LMC and SMC distances based on theoretical relations are
$18.51\pm0.07$ and $19.04\pm0.09$ respectively. On the other hand, the
same distances based on empirical relations are: $18.40\pm0.10$ and
$18.98\pm0.10$. This means that the two estimates agree at most within 5\%.
On the other hand, if we rely on the relations that are only marginally 
affected by reddening corrections, i.e. $PL_K$ and Wesenheit function, we 
find that the mean distances are: $18.53\pm0.08$-$19.04\pm0.11$ (theory) 
and $18.48\pm0.13$-$19.01\pm0.13$ (empirical). Predicted and empirical 
distance do agree at the level of 2\%.  

Previous findings support the use of a noncanonical ML relation,
and indeed theoretical predictions suggest that the zero-point of PL 
relations, in contrast with the slopes, do depend on the ML relation 
(Bono et al. 1999). It turns out that PL relations and $W$ function 
based on pulsation models that assume a canonical ML relation 
(Marconi et al. 2002, in preparation) give distances $\approx 0.14$
larger than previous ones. 
It is worth noting that predicted and empirical PL relations 
and Wesenheit function for FOs (Tables 2 and 3) do not depend, within 
the errors, on metal content. 

\section{Discussion and Conclusions} 

Figure~3 shows the comparison between theory and observations for LMC
FO variables. Note that to investigate in detail the accuracy of
current models we plotted the blue and the red edge of the instability
region instead of mean relations. Data plotted in this figure do
suggest that, within current uncertainties, predicted edges agree
quite well with empirical data. The agreement found in the $M_I-\log
P$ plane suggests that the discrepancy in the distance moduli based on
the $PL_I$ relation is due to a mild overestimate of individual
reddening corrections. The dashed lines display the mean relations
based on linear models provided by BA01. The comparison discloses 
that these relations
marginally account for empirical data, since they are located close to
blue edges of the instability strip. The difference between the two
sets of pulsation models is mainly due to fact that BA01 adopted a
canonical ML relation. Note that current FO edges do not show, 
in agreement with empirical data, a change in the slope when 
moving from 3.25 to 4 $M_\odot$ (Alibert et al. 1999).

The discrepancies we found are somehow at odds with the results
obtained by BA01 concerning the difference between the slopes of their
$W$ functions and the slopes predicted by Caputo, Marconi, Musella (2000)
for F pulsators.  The main difference is that BA01 were forced to 
extrapolate the $W$ function given by Caputo et al. (2000). In fact, 
these relations do rely on F models with periods ranging from 
$\log P = 0.50$ to $\log P=1.86$ for Z=0.004 and from
$\log P=0.51$ to $\log P=1.93$ for Z=0.008. However, both theoretical
predictions (BCM01) and empirical data (Bauer et al. 1999; G00) 
support the evidence that the slope of the F edges changes when 
moving from higher to lower luminosities. This means that
the extrapolation of PL and $W$ relations toward shorter periods is
risky, and therefore we did not perform any selection among unstable 
F and FO models.

The empirical lower limit in the period distribution of F and overtone
Cepheids is a key observable to constrain the accuracy of pulsational
and evolutionary models, since it supplies tight constraints on the
minimum mass whose blue loop crosses the instability strip, as well as
on the topology of the strip (Alcock et al. 1999; Bono et al. 2000; 
Beaulieu et al. 2001). On the other hand, the upper limit
in the period distribution of FO Cepheids is a robust observable to
constrain the plausibility of pulsation models (Bono et al. 1999;
FBK00). FOs located close to the {\em intersection point}, i.e. the
region of the instability strip where the blue and the red edge of FOs 
intersect, can be adopted to constrain the luminosity above which 
Cepheids only pulsate in the F mode. The longest predicted FO period 
is the aftermath of the topology of the instability strip, and in turn 
of the physical assumptions adopted to construct the pulsation models.
According to empirical evidence based on the OGLE database the longest
FO periods range from $\log P\approx 0.65$ to 0.77 for the SMC and
from $\log P\approx0.75$ to 0.80 for the LMC. Empirical estimates
based on different Cepheid samples give quite similar upper limits
for FOs (Beaulieu \& Marquette 2000). Note that previous upper limits only
bracket 4 (SMC) and 6 (LMC) FO Cepheids respectively. Moreover, the 4
SMC FOs with longer periods present a peculiar position in the
Wesenheit plane, and therefore they could have been misclassified.
These objects deserve a more detailed analysis to assess whether they
are genuine FO pulsators.  To avoid deceptive errors in the comparison
between theory and observations we decided to adopt 0.65 and 0.75 as 
upper limits in the period cut-off.

Current predictions suggest that the cut-off periods for FO Cepheids 
in MCs are located at $\log P\approx0.77$ (SMC) and $\log P\approx0.73$ 
(LMC). The predicted SMC upper limit is $\approx 0.1$ dex longer than 
observed, whereas the predicted LMC upper limit agrees quite well with 
the empirical one. Therefore it is not clear whether this discrepancy 
is due to a limit of theoretical models. It is noteworthy that pulsation 
models provided by BA01 predict stable FOs at periods longer than
$\log P\approx 1.3$ (SMC) and $\log P\approx 1.4$ (LMC). These upper
limits are substantially longer than observed ones. The reason for 
this difference is not clear. However, pulsation models based on  
a similar theoretical framework (linear, nonadiabatic),  and treatment 
of convective transport, predict similar cut-off periods (Chiosi et al. 1993).
To investigate the reasons of the mismatch between predicted and 
observed SMC FO cut-off periods we compared our predictions with 
empirical data for Galactic FO Cepheids. Empirical estimates 
suggest that the longest Galactic FO periods range from 
$\log P\approx 0.7$ to 0.88, but only 3 objects are included in
this period range (Kienzle et al. 1999). Current models at solar 
chemical composition predict a cut-off period, $\log P\approx 0.6$, 
that is $\approx 0.1$ dex shorter than the observed one.  
Note that nonlinear, convective models
constructed by FBK00 predict an upper limit that is 0.25 dex longer
($\log P\approx 0.95$) than observed. Since these models rely on a 
similar theoretical framework the difference could be due 
to the different treatment in the turbulent convection model 
(see \S 4.1 in FBK00) and/or to the adopted ML relation. Synthetic 
Color-Magnitude diagrams that account for evolutionary and pulsation 
properties are mandatory to constrain star formation rates and/or 
theoretical predictions (Alcock et al. 1999). 

The main conclusion we can draw is that current predictions concerning 
FO periods at the {\em intersection point} among Galactic and MC Cepheids 
agree within 0.1 dex with empirical values. A large sample of FO Cepheids 
in a different 
stellar system could be crucial to improve the accuracy of the empirical 
scenario. Accurate radius estimates of Galactic long-period FOs are 
strongly required to identify their pulsation mode. Different theoretical 
frameworks predict longer cut-off periods. This means that this observable 
can be adopted to validate the plausibility of pulsation models.

We thank D. Bersier as a referee for useful suggestions that improved 
the paper. This work was supported by MIUR/Cofin2000 under the project 
{\em Stellar Observables of Cosmological relevance}.

\pagebreak

\tablewidth{0pt}
\begin{deluxetable}{lcccc}
\tablecaption{Galactic relations of the form $M = a \times \log P + b$ 
using fixed theoretical slopes}\label{tbl-2}
\tablehead{
\colhead{$S$-$M$\tablenotemark{a}}&
\colhead{$a_{\rm theory}$}& 
\colhead{$b_{\rm theory}$} &
\colhead{$b_{\rm Hipparcos}$} &
\colhead{$N$\tablenotemark{b}} 
}
\startdata
\multicolumn{5}{c}{noncanonical models} \\
FO-$W$   & $-3.60\pm 0.04$ & $-3.12\pm 0.03$ & $-3.00\pm 0.14$ & 27 \\
FO-$V$   & $-3.01\pm 0.13$ & $-1.66\pm 0.09$ & $-1.65\pm 0.14$ & 31 \\
FO-$I$   & $-3.24\pm 0.09$ & $-2.23\pm 0.06$ & $-2.19\pm 0.16$ & 27 \\
FO-$K$   & $-3.52\pm 0.05$ & $-2.87\pm 0.03$ & $-2.86\pm 0.35$ & 7\\
F-$W$   & $-3.12\pm 0.04$ & $-2.79\pm 0.11$ & $-2.87\pm 0.16$ & 163 \\
F-$V$   & $-2.27\pm 0.12$ & $-1.37\pm 0.35$ & $-1.97\pm 0.15$ & 204 \\
F-$I$   & $-2.60\pm 0.09$ & $-1.93\pm 0.25$ & $-2.34\pm 0.16$ & 163 \\
F-$K$   & $-3.09\pm 0.04$ & $-2.52\pm 0.11$ & $-2.72\pm 0.19$ & 56 \\

\multicolumn{5}{c}{canonical models} \\
FO-$W$   & $-3.75\pm 0.04$ & $-3.26\pm 0.03$ & $-2.92\pm 0.14$ & 27 \\
FO-$V$   & $-3.25\pm 0.19$ & $-1.84\pm 0.15$ & $-1.52\pm 0.14$ & 31 \\
FO-$I$   & $-3.44\pm 0.13$ & $-2.40\pm 0.10$ & $-2.08\pm 0.16$ & 27 \\
FO-$K$   & $-3.65\pm 0.07$ & $-3.03\pm 0.05$ & $-2.80\pm 0.35$ & 7 \\
F-$W$   & $-3.03\pm 0.03$ & $-3.05\pm 0.10$ & $-2.95\pm 0.16$ &  163 \\
F-$V$   & $-1.98\pm 0.08$ & $-1.85\pm 0.25$ & $-2.25\pm 0.15$ &  204 \\
F-$I$   & $-2.39\pm 0.06$ & $-2.32\pm 0.19$ & $-2.53\pm 0.16$ &  163 \\
F-$K$   & $-3.01\pm 0.03$ & $-2.77\pm 0.10$ & $-2.78\pm 0.18$ & 56 \\

\enddata
\tablenotetext{a}{Predicted Wesenheit (W) and PL relations for First 
Overtone (FO) and Fundamental (F). ~$^b$ Number of objects used to 
derive the HIPPARCOS solution.
}
\end{deluxetable}

\begin{deluxetable}{lcccc}
\tablecaption{MC distance moduli obtained using predicted FO $PL$-relations.}\label{tbl-2}
\tablehead{
\colhead{$S$-$M$\tablenotemark{a}}&
\colhead{Slope$_{\rm T}$\tablenotemark{b}}&
\colhead{ZP$_{\rm T}$\tablenotemark{b}}&
\colhead{ZP$_{\rm E}$\tablenotemark{c}}&
\colhead{DM\tablenotemark{d}}
}
\startdata
\multicolumn{5}{c}{LMC} \\

1-$W$   & $-3.64 \pm 0.03$ & $-3.09 \pm 0.04$ & 15.45 $\pm$ 0.08 & 18.54$\pm$ 0.09 \\
2-$I_0$ & $-3.31 \pm 0.07$ & $-2.29 \pm 0.09$ & 16.13 $\pm$ 0.14 & 18.42$\pm$ 0.17 \\
3-$K_0$ & $-3.57 \pm 0.03$ & $-2.89 \pm 0.05$ & 15.62 $\pm$ 0.13 & 18.51$\pm$ 0.14 \\

\multicolumn{5}{c}{SMC} \\

4-$W$   & $-3.62 \pm 0.03$ & $-3.07 \pm 0.04$ & 15.99 $\pm$ 0.12 & 19.06$\pm$ 0.13\\
5-$I_0$ & $-3.27 \pm 0.07$ & $-2.31 \pm 0.09$ & 16.67 $\pm$ 0.19 & 18.98$\pm$ 0.21\\
6-$K_0$ & $-3.55 \pm 0.03$ & $-2.89 \pm 0.05$ & 16.13 $\pm$ 0.18 & 19.02$\pm$ 0.19\\

\enddata
\tablenotetext{a}{Solution. $^b$Predicted slope and Zero-Point (ZP). 
$^c$Empirical ZP, by fixing the slope to the theoretical one. 
Cuts are applied in period as follows: solutions 1 and 2 -- none, 
solutions 3, 4 and 5 -- $\log P > 0.25$, solution 6 -- $\log P > 0.30$.
$^d$Distance modulus.  
}
\end{deluxetable}

\begin{deluxetable}{lcccc}
\tablecaption{MC distance moduli obtained using empirical FO $PL$-relations.}\label{tbl-2}
\tablehead{
\colhead{$S$-$M$\tablenotemark{a}}&
\colhead{$a_{\rm obs.}$\tablenotemark{b}}& 
\colhead{$b_{\rm obs.}$\tablenotemark{b}} &
\colhead{$b_{\rm Hipparcos}$\tablenotemark{c}} &
\colhead{$DM$\tablenotemark{d}}
}

\startdata
\multicolumn{5}{c}{LMC} \\

1-$W$   &  $-3.40$ & $15.373 \pm 0.007$ & $-3.11 \pm$ 0.14 & 18.48 $\pm$ 0.14 \\
2-$I_0$ &  $-3.31$ & $16.134 \pm 0.001$ & $-2.15 \pm$ 0.16 & 18.28 $\pm$ 0.16 \\
3-$K_0$ &  $-3.39$ & $15.540 \pm 0.031$ & $-2.91 \pm$ 0.35 & 18.45 $\pm$ 0.35 \\

\multicolumn{5}{c}{SMC} \\

4-$W$   &  $-3.36$ & $15.881 \pm 0.033$ & $-3.14 \pm$ 0.14 & 19.02 $\pm$ 0.14 \\
5-$I_0$ &  $-2.87$ & $16.520 \pm 0.050$ & $-2.39 \pm$ 0.16 & 18.91 $\pm$ 0.17 \\
6-$K_0$ &  $-3.10$ & $15.936 \pm 0.068$ & $-3.03 \pm$ 0.35 & 18.97 $\pm$ 0.35 \\

\enddata
\tablenotetext{a}{Solution.
$^b$ Observed FO $PL$-relation; $W,K$ from G00, $I$-band 
derived here.  Cuts are applied in period as follows: solutions 1 and 2 -- 
none, solutions 3, 4 and 5 -- $\log P > 0.25$, solution 6 -- $\log P > 0.30$.
$^c$ Galactic zero-point based on HIPPARCOS data for the observed slope.  
$^d$ Distance modulus.
}
\end{deluxetable}

\pagebreak

\begin{figure}[ht]
\begin{minipage}{0.6\textwidth}
\resizebox{\hsize}{!}{\plotone{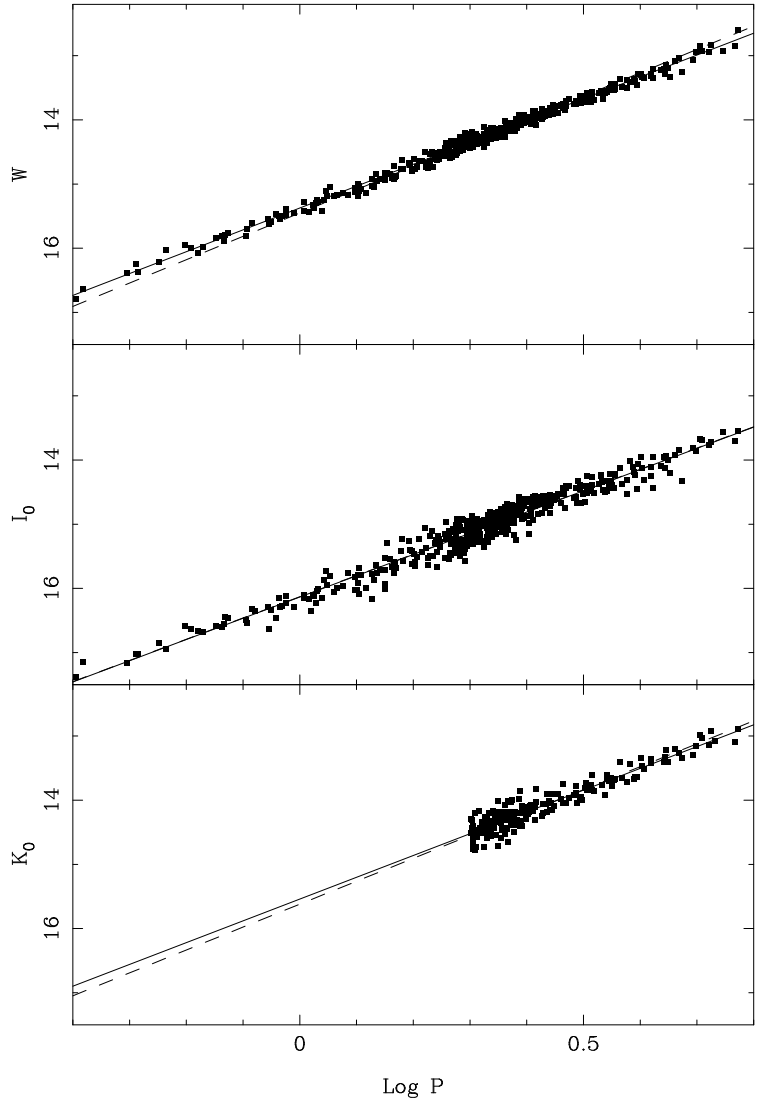}}
\end{minipage}
\caption{
LMC, $PL$-relation in Wesenheit function, $I_0$ and $K_0$ (from top to
bottom). Shown are the data (cut at $\log P = 0.3$ in $K$), the best
fitting linear relation (solid line), and the relation determined by
fixing the slope to its theoretical value.
}
\label{fig:lmc}
\end{figure}

\begin{figure}[ht]
\begin{minipage}{0.6\textwidth}
\resizebox{\hsize}{!}{\plotone{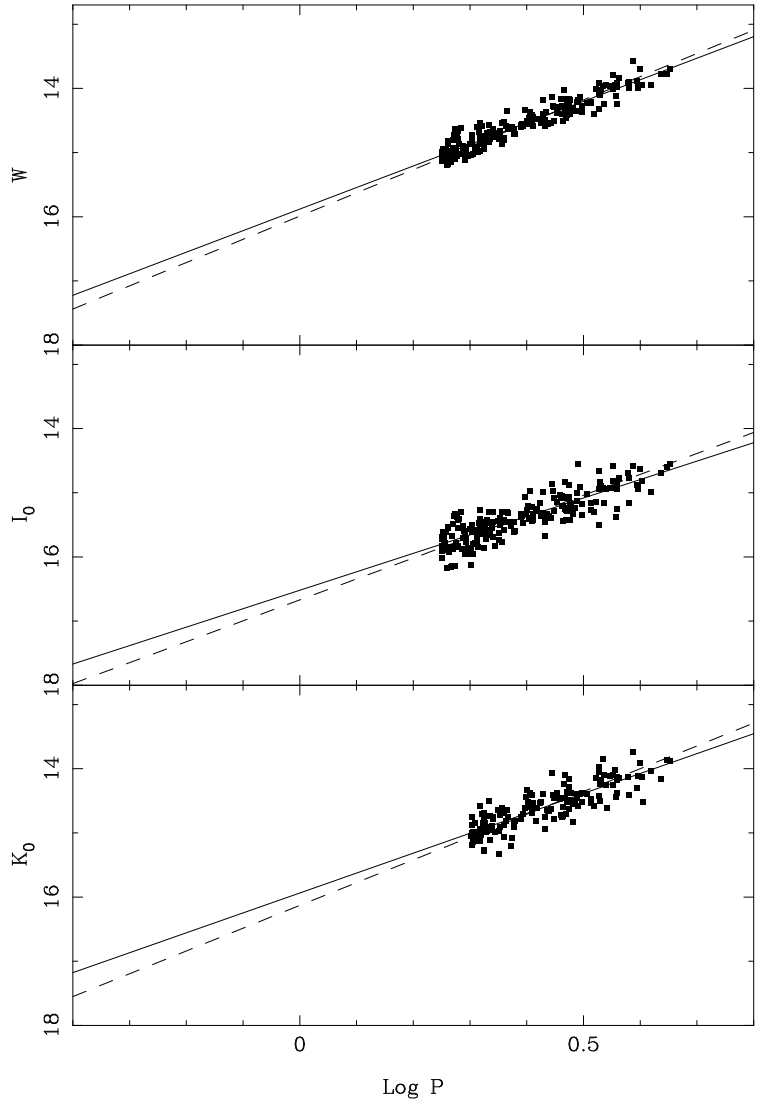}}
\end{minipage}
\caption{
Same as Fig. 1, but for SMC Cepheids. The cut is at $\log P = 0.3$ ($K$), 
and 0.25 ($W$, $I$).
}
\label{fig:smc}
\end{figure}

\begin{figure}[ht]
\begin{minipage}{0.8\textwidth}
\resizebox{\hsize}{!}{\plotone{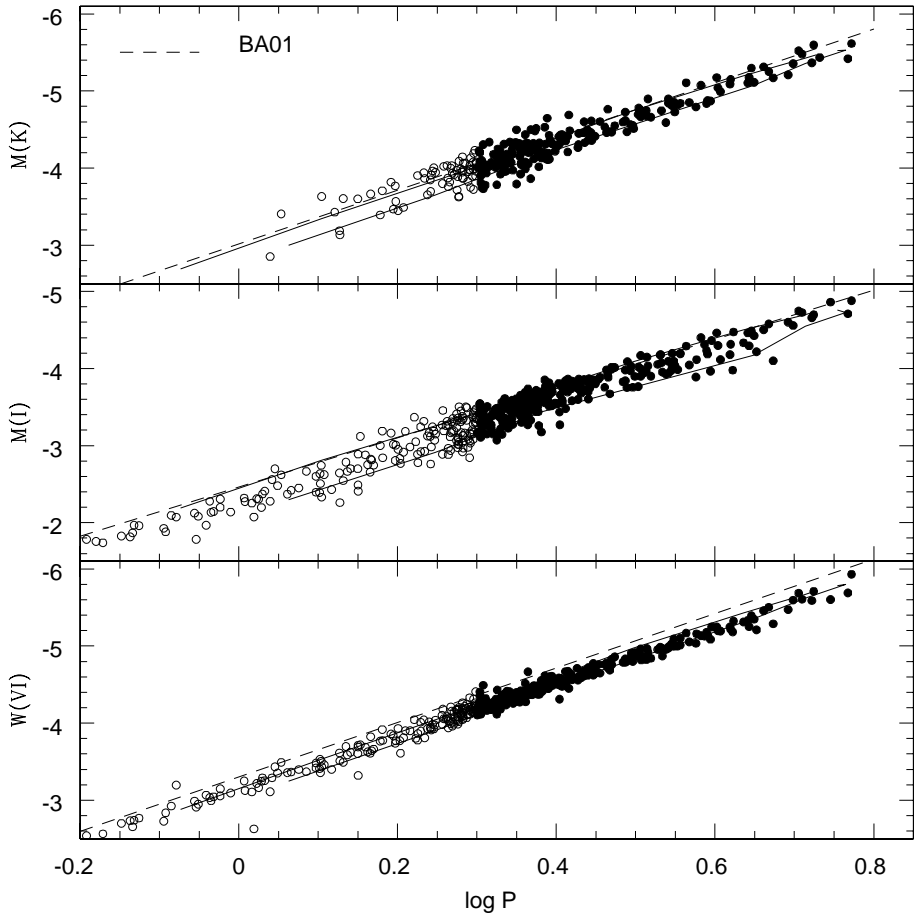}}
\end{minipage}
\caption{Comparison between predicted blue and red edges with
LMC empirical data. Filled symbols show Cepheids adopted to 
estimate the distance. Dashed lines display mean PL relations 
provided by BA01. Predicted edges were plotted using the 
distances listed in Table 2.}
\label{fig:strip}
\end{figure}

\end{document}